\documentclass[page-classic]{epl2}

\usepackage{amsmath} \usepackage{amssymb}
\usepackage{color}
\usepackage{graphicx}

\newcommand{\ket}[1]{\ensuremath{\left | #1 \right \rangle}} 
\newcommand{\bra}[1]{\ensuremath{\left \langle #1 \right |}} 
\newcommand{\mean}[1]{\ensuremath{\left \langle #1 \right \rangle}} 
\newcommand{\diff}{\ensuremath{\mathrm{d}}} 
\newcommand{\spvec}[1]{\ensuremath{\mathbf{#1}}} 

\newcommand{\FuncD}[1]{\ensuremath{\mathcal{D}#1}} 

\newcommand{\qop}[1]{\ensuremath{\hat{#1}}}

\newcommand{\EMscalar}[0]{\ensuremath{\varphi}} 
\newcommand{\vs}[0]{\ensuremath{v_{\text{S}}}}
\newcommand{\vsvec}[0]{\ensuremath{\spvec{\vs}}} 

\newcommand{\vcrit}[0]{\ensuremath{v_{\text{crit}}}}
 
\newcommand{\Ek}[1]{\ensuremath{E_{#1}}} 
\newcommand{\EkDeltaSq}[1]{\ensuremath{\Omega_{#1}}} 

\newcommand{\ephase}[0]{\ensuremath{\phi}}
 
\newcommand{\taumeasure}{\hbar} 
 
\newcommand{\nambuvec}[1]{\ensuremath{\boldsymbol{#1}}}
\newcommand{\nambumat}[1]{\ensuremath{\boldsymbol{#1}}} 

\newcommand{\GenFunc}[2]{\ensuremath{\mathcal{#1} \left [ #2 \right
]}}

\newcommand{\leftcurrket}[0]{\ensuremath{\ket{\circlearrowleft}}}
\newcommand{\rightcurrket}[0]{\ensuremath{\ket{\circlearrowright}}}
\newcommand{\leftcurrbra}[0]{\ensuremath{\bra{\circlearrowleft}}}
\newcommand{\rightcurrbra}[0]{\ensuremath{\bra{\circlearrowright}}}

\newcommand{\fermi}[0]{\ensuremath{\mu_{\text{F}}}}
\newcommand{\fermiv}[0]{\ensuremath{v_{\text{F}}}}

\newcommand{\Dmeanvs}[0]{\ensuremath{\delta\!\mean{\vsvec}}}

\newcommand{\Dmeanjpos}[1]{\ensuremath{\delta\!\mean{\spvec{j}(#1)}}}

\newcommand{\length}[0]{\ensuremath{L}}  

\newcommand{\DNtot}[0]{\ensuremath{\Delta N_{\text{tot}}}}

\newcommand{\Ip}[0]{\ensuremath{I_{\text{p}}}} 
\newcommand{\DIp}[0]{\ensuremath{\delta
\Ip}}

\title{The size of macroscopic superposition states in flux qubits}

\author{J. I. Korsbakken\inst{1,2} \and F. K. Wilhelm\inst{3} \and K. B. Whaley\inst{2}}
\shortauthor{J. I. Korsbakken \etal}

\institute{                    
  \inst{1} Department of Physics and Berkeley Quantum Information and Computation Center,
University of California, Berkeley, 94720, USA \\
  \inst{2} Department of Chemistry and Berkeley Quantum Information and Computation Center, University of California, Berkeley, CA 94720, USA \\
  \inst{3}  Institute for Quantum Computing and Department of Physics and
Astronomy, University of Waterloo, 200 University Avenue West, Waterloo, ON, Canada, N2L 3G1}

\abstract{The question as to whether or not quantum mechanics is applicable to the macroscopic scale has motivated efforts to generate superposition states of macroscopic numbers of particles and to determine their effective size.
Superpositions
of circulating current states in flux qubits constitute candidate states that
have been argued to be at least mesoscopic. 
We present a microscopic analysis that reveals the number of
electrons
participating in these superpositions
to be surprisingly but not trivially small, even though differences in macroscopic observables
are large.
}

\begin{document}

\maketitle

\section{Introduction}

Quantum mechanics makes predictions that often
seem at odds with the intuition gained through human experience at the macroscopic level.
Schr\"{o}dinger's cat paradox~\cite{Schroedinger35} provides an extreme example of this, highlighting the consequences of extrapolating basic quantum concepts such as superposition from microscopic
to macroscopic systems. 
The question of whether quantum behavior is restricted for large numbers of particles by some unknown non-quantum mechanism or contains some
limitation that we do not yet understand  is a fundamental unresolved question~\cite{Leggett2002}.  Realizing true macroscopic quantum superpositions would constitute one step towards an answer, providing evidence against macroscopic realism~\cite{Leggett85}.  

There are 
many different ways in which one could define the effective ``size'' of a macroscopic superposition state. One might first compare the absolute value of the difference in some suitable physical observable between the two branches to some characteristic atomic scale for
 that observable. However, as discussed in Ref.~\cite{Leggett2002}, this measure is not an unambiguous signature of a large quantum superposition of microscopic degrees of freedom.  
In order to understand how many microscopic degrees of freedom actively participate in the superposition, it is essential to further make a microscopic analysis of the differences between the two branches of a superposition state.  In this work we provide for the first time such a microscopic analysis for the superposition states made with flux qubits that show macroscopically distinct currents in different branches and which have consequently been suggested to involve superposition of macroscopic numbers of electrons~\cite{Leggett2005}.   

Superconducting flux qubits composed of micron-scale superconducting loops containing Josephson
 junctions and involving currents 
 attributable to $10^6$-$10^{10}$ 
electrons~\cite{Mooij99,WalHaarWilhelm2000,FriedmanPatelChen2000,Leggett2002} provide good candidates for forming macroscopic quantum superpositions.  Recent
experiments have demonstrated that
flux qubits
can be brought into a superposition of
macroscopically distinct states of circulating currents,
of the form $\leftcurrket + \rightcurrket$~\cite{WalHaarWilhelm2000,FriedmanPatelChen2000}, and that coherent oscillations between these states can be induced~\cite{Chiorescu03}.  
Here, the states $\leftcurrket$ and $\rightcurrket$ correspond to states of fluctuating persistent current with nonzero mean values in clockwise and anticlockwise directions around the loop, respectively~\cite{Orlando99}.
However, while the two components of this superposition state have macroscopically distinct values of magnetic flux and operation of flux qubits may be described in terms of a few macroscopic variables~\cite{Mooij99,WalHaarWilhelm2000,FriedmanPatelChen2000,Chiorescu03,Insight},
it is not clear how many \textit{microscopic} degrees of freedom participate in the superposition~\cite{Leggett2002,Leggett2005,MarquardtAbelDelft2008}.
Answering this fundamental question requires a fully microscopic analysis for the number of electrons participating in the superposition, a challenging task which has hitherto not been undertaken for superconductors.
We present here a first principles approach that allows us to place a bound on this size and evaluate this for experimental flux qubit systems.  
We find that for all flux qubits fabricated to date, this results in a  superposition state whose size is considerably smaller than the number of electrons carrying the supercurrent.  Our microscopic analysis reveals that the indistinguishability and fermion statistics of the constituent Cooper pair electrons play a critical role in this reduction.  

Our calculations are made at $T=0$ and are based on the framework of BCS theory~\cite{BardeenCooperSchrieffer1957}, in which pairing is described by a complex-valued
order parameter $\Delta(\vec{r})=|\Delta|e^{i\ephase}$. In order to describe a flux qubit, we use the path integral description~\cite{EckernSchonAmbegaokar1984} of a BCS superconductor in a ring geometry, interrupted by small capacitance Josephson junctions. In this geometry, 
Coulomb repulsion at the junction drives quantum fluctuations of $\ephase(\vec{r})$.
Cat states in flux qubits are described as superpositions of many-body states with clockwise and counterclockwise net circulating currents, which we denote by $\frac{1}{\sqrt{2}} \left ( \leftcurrket + \rightcurrket \right)$.  In each of the two branches of the superposition, the phase is given by a narrow distribution with mean value $\phi_{L/R}(\vec{r})$, respectively. For qubits with thickness much smaller than the penetration depth $\lambda$, we can assume that the current flow is uniform in the
lateral dimension (the thickness, or vertical dimension of a planar geometry loop with horizontal orientation).  The order parameter and the vector potential $\vec{A}$ together define the \emph{superfluid velocity} $\vsvec
\equiv - \frac{\hbar}{2m} \left ( \nabla \phi + \frac{2e}{\hbar c}
  \spvec{A} \right )$, with $m$ the single electron mass and $e$ the single electron charge. $\vsvec$ is equal to the mean velocity of
superconducting electrons in the system as normally defined in terms of the center of mass velocity of Cooper pairs and is related to
the electron current density via $\spvec{j} = e \rho_e \vsvec$, with
$\rho_e$ an effective electron density~\cite{Schmidt97}.  (The Cooper pair superfluid density is equal to $\rho_e/2$.)

As long as
both the externally applied and the internally generated magnetic
fields are weak enough and the cooling is sufficiently
adiabatic to avoid vortex formation,
we can decompose the entire electron
system in terms of single-electron modes labeled by the linear momentum $\spvec{p} \equiv \hbar \spvec{k}$ and spin~$\sigma$, where $\spvec{k}$ is the eigenvalue of $-i\nabla$. To analyze the current carrying states, we define creation and annihilation operators $\qop{c}_{\spvec{q},\sigma}^{\dag}$ and $\qop{c}_{\spvec{q},\sigma}$, where $\spvec{q}$ is the eigenvalue of  $-i\nabla + \frac{m}{\hbar} \, \vsvec$.
The index $\spvec{k}$ labels the internal momentum of Cooper pairs.  
In BCS theory, the average occupation number of a single-electron mode is given by $n_{\spvec{k}} =\mean{\qop{c}_{\spvec{k},\sigma}^{\dag} \qop{c}_{\spvec{k},\sigma}}$ ($n_{\spvec{k}}$ is independent of $\sigma$ for any realistic magnetic field strength) and at zero supercurrent is equal to $n_k=(1-E_k/\Omega_k)/2$ with
$ \Ek{\spvec{k}} = \frac{\hbar^2 k^2}{2m} -\fermi$, 
$\EkDeltaSq{\spvec{k}} =\sqrt{\Ek{\spvec{k}}^2 +\Delta^2}$ and 
$\fermi$ the Fermi energy, while the amplitude of the Cooper pair mode (the 'condensation amplitude'~\cite{deGennes66}) 
$\mean{\qop{C}_{\spvec{k}}^{\dag}} \equiv \mean{\qop{c}_{\spvec{k},\uparrow}^{\dag} \qop{c}_{-\spvec{k},\downarrow}^{\dag}}$ 
has a smooth peak of width $\simeq \Delta/v_F$ at the Fermi momentum $k_F$~\cite{deGennes66}.
We emphasize that the circulating current is a dissipationless supercurrent.  The modes $\spvec{k}$ and $\spvec{q}$
should not be confused with quasiparticle states that carry {\em dissipative} processes and 
that 
are separated from the condensate by an energy gap.

\section{Microscopic Analysis of Flux Qubit Superposition Size}
We make here a microscopic, all-electron analysis of the superposition size using the distinguishability measure of Ref.~\cite{KorsbakkenWhaleyDubois2007}.   The latter provides an operational measure that asks what is the largest number of elementary constituents in an $N$ particle system that need to be measured, within a given precision, to collapse the superposition to a single branch.  Ref.~\cite{KorsbakkenWhaleyDubois2007} used this to show that the number of constituents that participate actively in the superposition, i.e., that do something different in the two branches, is given by $N/n$, where $n$ is the smallest 
number of particles that must be measured to guarantee successfully 
distinguishing between the two branches. 
For flux qubit states. the electron indistinguishability must additionally be taken into account, which is done by extending the distinguishable particle  analysis of Ref.~\cite{KorsbakkenWhaleyDubois2007} to measurements of electron mode occupation, with the density matrices replaced by fermionic multi-mode density matrices~\cite{Korsbakken2008}.  The microscopic number of constituents (modes) actively participating in the superposition is then determined by the number of 
$n$-electron modes that are required to be measured in the $N$ mode electron system to successfully distinguish the two branches. Expanding the reduced mode density matrices in the plane wave basis $\spvec{q}$, we find that for small differences in mode occupancy between $L$ and $R$ branches, the limiting number of mode measurements can be bounded by $N / \DNtot$ where $\DNtot$ is the total difference in mode occupation number between the two branches~\cite{Korsbakken2009}, leading to a bound $\DNtot$ on the superposition size. This simple-looking result is underscored by a more profound fact, namely that when the indistinguishability of electrons is taken into account, $\Delta N_{tot}$ is the only meaningful indicator of how many electrons are affected when passing from one
branch of the superposition to the other.  

It can be shown that this choice of single-particle basis maximizes the occupation number differences between the two branches,
$\delta n_{\spvec{q}} = \leftcurrbra \qop{c}_{\spvec{q},\sigma}^{\dag} \qop{c}_{\spvec{q},\sigma} \leftcurrket \, - \, \rightcurrbra \qop{c}_{\spvec{q},\sigma}^{\dag} \qop{c}_{\spvec{q},\sigma} \rightcurrket$~\cite{Korsbakken2009}, thereby ensuring that $\DNtot$ is a mode-independent and hence true upper bound on the cat size.  We show below that this calculation of superposition size in terms of single electron modes can be directly related to the corresponding calculation in terms of Cooper pair modes.  While the latter appears more natural, it requires more complex calculations and, as we discuss below, for small superfluid velocities $\vsvec$, gives identical results to calculations based on single electron modes.   

$\Delta N_{tot}$ can be calculated microscopically in the path integral approach within the BCS description.
The average occupation number of a single-electron mode is given by $n_{\spvec{q}} =\mean{\qop{c}_{\spvec{q},\sigma}^{\dag} \qop{c}_{\spvec{q},\sigma}}$ (as noted above, $n_{\spvec{q}}$ is independent of $\sigma$ for any realistic magnetic field strength). 
The necessary expectation values  can be obtained from the generating functional
\begin{equation}
\begin{split}
\GenFunc{Z}{\xi} \,  = \, \int \FuncD{\psi} \, \FuncD{\spvec{A}} \, \FuncD{\EMscalar} \, &
e^{-S[\psi,\spvec{A},\EMscalar]
+\sum_{\sigma} \int_0^{\hbar\beta} \frac{\diff \tau}{\taumeasure} \, \int \diff^3 r \, \left [ \xi_{\sigma}^*(\spvec{r},\tau) \psi(\spvec{r},\tau) + \xi_{\sigma}(\spvec{r},\tau) \psi_{\sigma}^*(\spvec{r},\tau) \right ]},
\end{split}
\label{eq:GeneratingFunctional1}
\end{equation}
where $\psi (\psi^{\dagger})$ are the electron field operators, $\spvec{A}$ the vector potential, 
$\EMscalar$ the electromagnetic scalar potential and $S[\psi,\spvec{A},\EMscalar]$ the action functional of Ref.~\cite{EckernSchonAmbegaokar1984}.  Employing a Hubbard-Stratonovich transformation and saddle point analysis as in that work introduces the order parameter  $\Delta(\vec{r})$ and reduces the generating functional to a form in which only the phase $\ephase$ remains as a non-classical variable, namely,
\begin{equation}
\GenFunc{Z}{\nambuvec{\xi}} \, = \, \int \FuncD{\ephase} \, e^{- \int_{0}^{\hbar\beta} \frac{\diff\tau}{\hbar} \left [ \frac{1}{2} C \dot{\Phi}^2 - E_J \, \cos \frac{2\pi \Phi}{\Phi_0} + \frac{1}{2L} \left ( \Phi - \Phi_{\text{ext}} \right )^2 + \int \diff^3 r \, \nambuvec{\xi}^{\dag} \nambumat{G} \nambuvec{\xi} \right ]}.
\end{equation}
Here the flux $\Phi$ is related to the phase variable $\ephase$ through the  flux quantization relation
$\Phi = \frac{\ephase}{2\pi} \, \Phi_0$, the Josephson energy $E_J$ is an effective quantity 
~\cite{AmbegaokarBaratoff1963} and 
$\nambumat{G}$ is the sum of the zeroth order bulk Green's function and perturbative contributions from the superfluid flow and junction tunneling,
$\nambumat{G} \, = \left ( \nambumat{G}_{\text{bulk}}^{-1} + \nambumat{T_{rr'}} \right )^{-1} 
=  \, \nambumat{G}_0 + \delta\nambumat{G}_{\vsvec} + \delta\nambumat{G}_T$.   
Standard analysis then leads to the single electron and Cooper pair mode correlation functions in terms of matrix elements of $\nambumat{G}$~\cite{AbrikosovGorkovDzyaloshinski1975}.  We analyze first the loop current contribution $\nambumat{G}_0 + \delta\nambumat{G}_{\vsvec}$ and discuss the tunneling contribution $\delta\nambumat{G}_T$ separately below.  The loop current contribution to the single mode correlation functions is obtained as
\begin{equation}
\begin{split}
\bra{\Psi} \qop{c}_{\spvec{q}\sigma}^{\dag} \qop{c}_{\spvec{q}\sigma} \ket{\Psi} \, &= \,
- \hbar \lim_{\tau\rightarrow 0^+} G_{\sigma\sigma}(\spvec{q}-\frac{m\vsvec}{\hbar},0;\spvec{q}-\frac{m\vsvec}{\hbar},\tau),
\end{split}
\label{eq:QBranchOccupation1}
\end{equation}
where $G_{\sigma\sigma}$ is a matrix element of $\nambumat{G}_0 + \delta\nambumat{G}_{\vsvec}$ indexed by the electron spin.
As long as the current in the superconductor is small compared to the critical current $I_{\rm c,bulk}$, the superfluid velocity $\vsvec$ will be a small perturbative quantity and the Green's functions can be expanded to 
first order in $|\vsvec|/\vcrit = |\vsvec|\fermiv m/ \Delta$, where $\fermiv$ is the Fermi velocity. 
Carrying this out yields~\cite{Korsbakken2009,Korsbakken2008}
 \begin{equation}
\begin{split}
\bra{\Psi} \qop{c}_{\spvec{q}\sigma}^{\dag} \qop{c}_{\spvec{q}\sigma} \ket{\Psi} \, &= \,
 \frac{1}{2} 
\left ( 1 - \frac{\Ek{q}}{\EkDeltaSq{q}} \right ) + \frac{1}{2} 
\frac{\Delta^2}{\EkDeltaSq{q}^3} \, \hbar \spvec{q} \cdot \mean{\vsvec},
\end{split}
\label{eq:QBranchOccupation2}
\end{equation}
where $\mean{\vsvec}$ is the mean superfluid velocity averaged over the quantum state of the system.
To first order in $\vsvec$,
the  difference in occupation number of a mode $(\spvec{q},\sigma)$ 
 between the two circulating current superposition branches,
 is then given by 
 \begin{equation}
 \delta n_{\spvec{q}} \, \equiv \, \leftcurrbra \qop{c}_{\spvec{q},\sigma}^{\dag} \qop{c}_{\spvec{q},\sigma} \leftcurrket \, - \, \leftcurrbra  \qop{c}_{\spvec{q},\sigma}^{\dag} \qop{c}_{\spvec{q},\sigma} \rightcurrket \, = \, \frac{\Delta^2}{2\EkDeltaSq{\spvec{q}}^3} \, \hbar \spvec{q} \cdot \Dmeanvs,
 \label{eq:dn_q}
 \end{equation}
 where $\Dmeanvs$ is the difference in superfluid velocity between the two branches.
  
 Analysis of the corresponding Cooper pair correlation functions 
\begin{equation}
\delta\mathcal{N}_{\spvec{k,-k}} = \, \leftcurrbra  \qop{c}_{\spvec{k}\uparrow}^{\dag} \qop{c}_{-\spvec{k}\downarrow}^{\dag} \qop{c}_{-\spvec{k}\downarrow} \qop{c}_{\spvec{k}\uparrow} \leftcurrket - \leftcurrbra \qop{c}_{\spvec{k}\uparrow}^{\dag} \qop{c}_{-\spvec{k}\downarrow}^{\dag} \qop{c}_{-\spvec{k}\downarrow} \qop{c}_{\spvec{k}\uparrow} \leftcurrket
\end{equation}
in the laboratory frame is complicated by the fact that the modes $\spvec{q} = \spvec{k} + m\vsvec/\hbar$ and $\spvec{q}' = \spvec{k}' + m\vsvec/\hbar$ will be coupled only if $\spvec{k}' = -\spvec{k}$, which means that whether two modes $\spvec{q}$ and $\spvec{q}'$ are Cooper pair coupled or not depends on the superfluid velocity.
Detailed calculation shows that if the modes are not Cooper pair correlated, there is no first order contribution to the correlation function $\delta\mathcal{N}_{\spvec{q},\spvec{q}'}$, while if they are Cooper pair correlated, the corresponding occupation number $\mathcal{N}_{\spvec{k}} \equiv \mean{\qop{C}_{\spvec{k}}^{\dag} \qop{C}_{\spvec{k}}} = \mean{\qop{c}_{\spvec{k},\uparrow}^{\dag} \qop{c}_{-\spvec{k},\downarrow}^{\dag} \qop{c}_{-\spvec{k},\downarrow} \qop{c}_{\spvec{k},\uparrow}}$ of a Cooper-pair mode is \emph{identical} to $n_{\spvec{k}}$~\cite{Korsbakken2008}. Hence, looking at Cooper-pair modes rather than single-electron modes does not change the effective cat size and from now on we shall consider just the single-electron quantity $\delta n_{\spvec{q}}$, eq.~(\ref{eq:dn_q}).

\section{Superposition Size Estimates}
The change of the mode occupation  $\delta n_{\spvec{q}}$ reflects the fact that a supercurrent gives an
additional momentum $2q$ to each Cooper pair, thereby changing the momentum 
structure of the condensate 
without exciting quasiparticles.  To obtain numerical estimates, we proceed by first determining the number of electrons changing modes \emph{per unit volume}, $\delta n(\spvec{r})$, at a point $\spvec{r}$ in the superconductor,
by integrating the absolute value of $\delta n_{\spvec{q}}$ over all $\spvec{q}$ values.  
(At $T=0$, there is no contribution from quasiparticle excitations.)
We can similarly obtain an expression for $\delta\spvec{j}(r)$, the difference in local current density,
 as an integral over $\spvec{q}$.  These two integrals can be simplified by noting that  the contribution of modes far from the Fermi surface is strongly suppressed in eq.~(\ref{eq:QBranchOccupation2}) and hence also in $\delta n_{\spvec{q}}$.
As shown in Ref.~\cite{Korsbakken2009}, this leads to the relation  
$\delta n(\spvec{r}) \, = \, 3\, \left | \Dmeanjpos{\spvec{r}} \right | / 4 \, e \fermiv$, 
where we have divided by $2$ in order to avoid double-counting electrons when they are removed from one mode and added to another. Integrating this expression
over the entire volume of the superconductor then gives the total difference in occupation numbers as
\begin{equation}
\DNtot \, = \, \frac{3\length}{4\, e \fermiv}
\, \DIp,
\end{equation}
where $\length$ is the total length of the main superconducting loop of the flux qubit and $\DIp$ the difference in ``persistent current''~\cite{WalHaarWilhelm2000} between the superposed branches.
The experimental parameters and corresponding values of $\DNtot$ for
the three recent experiment~\cite{FriedmanPatelChen2000,WalHaarWilhelm2000,HimeReichardtPlourde2006} are listed in table~\ref{tbl:ExperimentsAndEffectiveCatSizes}.
The SUNY experiment~\cite{FriedmanPatelChen2000} was carried out using a single-junction RF-SQUID configuration creating a superposition of two nearly degenerate highly excited states, while the Delft~\cite{WalHaarWilhelm2000} and Berkeley~\cite{HimeReichardtPlourde2006} experiments were both made with three-junction flux qubits that generated a superposition of degenerate ground states.
We also list in table~\ref{tbl:ExperimentsAndEffectiveCatSizes} the corresponding values for the difference in the two macroscopic observables current and magnetic moment between the two branches,  respectively $\delta I_p$ and $\delta \mu=A\delta I_p$, where $A$ is the area enclosed by the superconducting loop.  

\begin{table}
\caption{Parameters and effective cat sizes for current superposition states produced at SUNY~\cite{FriedmanPatelChen2000}, Delft~\cite{WalHaarWilhelm2000} and Berkeley~\cite{HimeReichardtPlourde2006}.
$\fermiv$ is the Fermi velocity, $L$ the length of the superconducting loop, $\delta I_p$ the measured difference in persistent current between the two branches and $\delta \mu = A \delta I_p$ is the difference in magnetic moment, where $A$ is the area enclosed by the loop.  $\DNtot$ is the upper bound microscopic estimate of effective superposition size, i.e., the total number of electrons participating in the superposition state.}
\label{tbl:ExperimentsAndEffectiveCatSizes}
\begin{center}
\begin{tabular}{ | c | c | c | c | c | c | c |}
\hline
Exp.  &  Mat.  &  $\fermiv [\un{m/s}]$  &  $L [\un{\mu m}]$  &  $\DIp$  &  $\Delta \mu/\mu_B$ &\DNtot  \\ \hline
SUNY  &  Nb  &  $1.37 \cdot 10^6 $  &  $560$  &  $2$--$3\un{\mu A}$  & $5.5-8.3 \cdot 10^9$ & $3800$--$5750$\\
Delft  &  Al  &  $2.02 \cdot 10^6 $  &  20 &  $900\un{nA}$  & $2.4\cdot 10^6$ & $42$  \\
UCB  &  Al  &  $2.02 \cdot 10^6 $  &  $183 $  &  $292 \un{nA}$  & $4.23\cdot 10^{7}$ & $124$  \\
\hline
\end{tabular}
\end{center}
\end{table}

Electron tunneling through the junction can also contribute to the difference in mode occupation in the two branches of the superposition.  We have estimated this contribution assuming momentum
conserving tunneling events~\cite{Korsbakken2008,Korsbakken2009} and find that the tunneling dynamics of electrons close to the junction do not significantly alter the superposition size estimates given here.
We have also analyzed the effects of impurities and defects, i.e., a 'dirty superconductor' on the robustness of these flux qubit superpositions.  For the high quality nanofabricated flux qubit circuits, these effects are restricted to elastic scattering of Cooper pairs from impurities.  Detailed analysis of the Green's functions given elsewhere~\cite{Korsbakken2009} shows that $\delta n_{\spvec{q}}$ is actually independent of the impurity concentration, implying that the results obtained here apply both to clean and dirty superconducting flux loops.

\section{Discussion}
The results in table~\ref{tbl:ExperimentsAndEffectiveCatSizes} show that while not trivially small, the number of electrons participating in the superpositions 
is considerably smaller than 
previous estimates that simply counted all electrons within a London penetration depth of the surface~\cite{FriedmanPatelChen2000,WalHaarWilhelm2000,Leggett2002,Leggett2005}.  
Furthermore, there is a marked contrast between the fact that the values of $\Delta N$ are well below what is generally regarded as macroscopic and the fact that the two branches nevertheless have macroscopically distinct values of
current and magnetic moment.  This apparent conundrum arises 
because
the electrons are circulating in opposite directions at high speeds ($\fermiv$) around a path enclosing a relatively large surface area. Our result shows that the actual number of electrons that would be found to be behaving
differently in the two branches if one could measure them at the microscopic
level is, however,
quite modest.
This discrepancy between a large difference in the value of an observable quantity 
(current $\delta I_p$ or magnetic moment $\delta \mu$)
and the small number of particles actively involved in the superposition, $\Delta N_{tot}$, derives from the indistinguishability and the fermionic character of electrons. 
Thus, given the indistinguishability of electrons, $\Delta N_{tot}$, i.e., the total change in occupation numbers of all electron modes in the system, is indeed the {\em only} meaningful
indicator of how many electrons are affected when passing from one
branch of the superposition to the other.  In addition, because of the fermion statistics, single electron modes (and also Cooper pair modes) corresponding to opposite momenta in the laboratory frame are located at diametric extremes on the Fermi surfaces for the two branches.  There is a large velocity difference between these electrons because $\fermiv \gg \vs$, even though the number of electrons involved is small. 

We now consider whether the superposition size could be increased to truly macroscopic values.  One option is to simply increase the number of modes available, by scaling up the physical dimensions of the system. Experimental efforts in this direction are underway 
~\cite{Mooij09}.  Another option is to build loops from small cross-section superconducting wires together with large area Josephson junctions, for which the larger current differences between the superposition branches may be possible~\cite{Korsbakken2009}.

The microscopic sizes calculated here for flux superpositions are generally larger than typical experimentally realized superposition 
states in molecular and optical systems~\cite{ArndtNairzVos-Andreae1999,Haroche08}. An intriguing aspect of the numerical estimates is that they are neither trivially small, i.e., not unity as expected from 
a quantum circuit analysis in terms of macroscopic variables~\cite{MarquardtAbelDelft2008}, nor truly macroscopic.  This placement between a microscopic and a macroscopic number of participating constituents reflects the key role of the fermi statistics of the electrons that emerges from our analysis.
Our analysis shows explicitly
that for fermionic systems it is quite possible to have
superpositions which appear large when judged by the magnitude of
physical observables, but which nevertheless involve only a relatively
small number of microscopic particles in the superposition. 
Somewhat surprisingly, this critical role of the particle statistics in differentiating these two fundamental properties of 
large scale superposition states has not been appreciated before. 
Thus, while there is no intrinsic size or number scale limiting the existence of macroscopic quantum superpositions, the present work shows that the quantum statistics of the constituent particles can nevertheless
be important for evaluating
the effective number of particles participating in a superposition whose branches are characterized by macroscopically distinct observables.  Whether the corresponding gap between difference in observable quantities and number of particles involved is smaller for superposition states in a bosonic system is an interesting question for further work. 
More generally, these results imply that existing superconducting flux qubit experiments 
provide realization of quantum superpositions on a
scale of a few thousand participating microscopic constituent particles. While large in comparison to other experimentally realized superpositions, they are not truly macroscopic and thus the formation of superposition states in which a macroscopic number of microscopic ``elemental" constituents behave differently in the two branches remains an open challenge.  

\acknowledgments
This work was supported by NSF through the ITR program and by NSERC through the
 discovery grants program.


\end{document}